\documentclass{article}

\usepackage{authblk}
\usepackage{url}
\usepackage{pdfpages}
\usepackage{float}
\usepackage{natbib}
\usepackage{amsthm}

\begin{document}
\title{The Minimum Hybrid Contract (MHC): Combining Legal and Blockchain Smart Contracts}

\author[1]{Jørgen Svennevik Notland}
\affil[1]{School of Entrepreneurship, NTNU, Trondheim, Norway\\jorgen.notland@protonmail.ch}
\author[2]{Jakob Svennevik Notland}
\affil[2]{Department of Computer Science, NTNU, Trondheim, Norway\\ijakob\_sn\_@hotmail.com}
\author[3]{Donn Morrison}
\affil[3]{Department of Computer Science, NTNU, Trondheim, Norway\\donn.morrison@ntnu.no}

\date{}




\maketitle

\begin{abstract}

Corruption is a major global financial problem with billions of dollars rendered lost or unaccountable annually. Corruption through contract fraud is often conducted by withholding and/or altering financial information. When such scandals are investigated by authorities, financial and legal documents are usually altered to conceal the paper trail. 

Smart contracts have emerged in recent years and appear promising for applications such as legal contracts where transparency is critical and of public interest. Transparency and auditability are inherent because smart contracts execute operations on the blockchain, a distributed public ledger.

In this paper, we propose the Minimum Hybrid Contract (MHC), with the aim of introducing 1) auditability, 2) transparency, and 3) immutability to the contract's financial transactions. The MHC comprises an online smart contract and an offline traditional legal contract. where the two are immutably linked. 

Secure peer-to-peer financial transactions, transparency, and cost accounting are automated by the smart contract, and legal issues or disputes are carried out by civil courts. The reliance on established legal processes facilitates an appropriate adoption of smart contracts in traditional contracts.

\end{abstract}

\section{Introduction}
The type of financial institutions a country ought to implement has, for the last three decades, mostly revolved around discovering and implementing the ideal mechanisms for achieving a free-market economy. At the heart of these debates has been the question of how a free market economy can be both secure, accountable, transparent, and inclusive to the actors involved, with some of the key solutions to these issues being particular financial regulations and legal contracts. Moreover, while it is the corruption occurring in developing and emerging economies outside the West, which gains most of the media attention, it is essential to note that the issue of corruption is also a significant challenge for developed countries. 

To overcome the challenges of corruption and reliance on a trusted centralized third party, we propose supplementing the existing legal regime with smart contracts where a smart contract supplements the legal contract. Our approach, called the Minimum Hybrid Contract (MHC), seeks to incorporate the accountability and transparency of smart contracts into the realm of human judgment and discretion.

An open-source and open access blockchain (hereby referred to as ``blockchain``) is a distributed database using a standard protocol such as Bitcoin. It is a dynamically updated ledger of records that is accessible for anyone that has an Internet connection. This paper demonstrates that using blockchain technology as a supplement can improve legal contracts. In agency theory, ``moral hazard,`` and ``information asymmetry`` -which are explained in detail below- are defined as recurrent issues in contractual relationships, which this paper sets out to overcome using blockchain technology.

The argument advanced here is that in addition to using a court for legal disputes in contractual relationships, supplementing the legal contract with blockchain as a tool for setting the rules for financial transactions provides transparency, immutable transactions while also lowering the cost of conducting due diligence and suitability assessments by lowering the cost of information sharing.

Section 3 presents a conceptual and software-architectural overview of an MHC. It consists of a legal contract that is supplemented by a smart contract. The smart contract facilitates transparent and secure transactions with immutable records, and a civil court handles disputes at the mercy of a legal contract between a principal and an agent.

The MHC is discussed in Section 4 as a tool for solving principal-agent problems and solving the issue of low trust in agent-principal relationships. The paper focuses on how an MHC is structured and discussion on how an MHC affects the principal-agent relationship. It is argued to provide a deterrent against financial crime. 

\section{Background and related work}

There have been several recent works porting traditional contracts onto blockchain technology to remove the need for a trusted central authority and/or provide transparency and auditability in contractual agreements \citep{weber2016untrusted,governatori2018legal,molina2018implementation}.

Weber et al. (2016) \citep{weber2016untrusted} propose a hybrid on- and off-blockchain technique integrating smart contracts into multi-party, collaborative business processes. Lack of trust and reliance on a central authority are central issues the approach aims to address. The on-blockchain component comprises the process instance contract, a smart contract, which is coupled with the off-blockchain component through a set of triggers (programs acting as an interface between on- and off-blockchain). While sharing some architectural similarities, the MHC we propose is different due to the way the smart and legal contracts are linked; the digital fingerprint of the legal document is embedded in the smart contract running on the blockchain. Furthermore, we do not replace legal contracts, but rather supplement them with a smart contract to allow transparency and auditability.

Governatori et al. (2018) \citep{governatori2018legal} focus on smart contracts with the absence of a traditional written legal contract. In other words, the smart contract is the legal contract, and its execution on the blockchain implements the terms of the contract, provides auditing, facilitates financial transactions, etc. The main contribution of the work examines the conceptual connection between legal and smart contracts and studies the suitability of imperative (for example, Solidarity) versus declarative languages for smart contracts.

Molina-Jimenez et al. (2018) \citep{molina2018implementation} argue for a hybrid on- and off-blockchain approach to contractual agreements. The authors implement a hybrid approach, but the reliance on a trusted third party remains in the off-blockchain component of contract enforcement. This is similar to the off-blockchain process we propose for the MHC, where courts handle legal contract disputes.

In the Centrifuge white paper by Vogelsang et al. \citep{Centrifuge}, a method is proposed for non-fungible tokens (NFTs) where on-blockchain tokens refer to off-blockchain documents. As will be apparent below, the MHC method we propose is different as it embeds the off-blockchain document fingerprint in the on-blockchain metadata, thereby immutably linking the two.

The evidence standard Ethereum improvement proposal number 1497\citep{EIP2018} mentions a possible reference from a smart contract to a legal contract. This proposal describes standards for using the Ethereum blockchain to store information on smart contracts immutably and used it for future reference in case of a court case so that arbitrators can accurately and fairly evaluate it.

\subsection{Contract fraud and corruption}

While the examples of corruption in the developed world mentioned thus far have been limited to the financial sector, it is essential to note that this issue also blights other sectors. For example, fraud in the construction industry is notable and takes the form of withholding information, misleading and altering documents. Other malpractices, such as making payments and invoices for materials never received is also prevalent  \citep{Heuvel2005, Bowen2007}. For example, two surveys conducted in South Africa and Australia have shown that deceit and misinformation are the most common types of fraud \citep{VeeSkitmore2003}.

Throughout several developed states in Europe, Danske Bank was involved in facilitating the biggest money-laundering scheme in human history involving \$230 billion in 2019 \citep{MilneWinter2018}. The world is experiencing a ``trust crisis'', where public trust in government and corporations reached a historic low in 2017 \citep{edelman2017} and remains relatively low in most of the world in 2019 \citep{Edelman2019}. In a global survey on trust in 2017, Edelman reports that 85\% of respondents indicated that they do not ``trust the system''. Moreover, only 52\% of the respondents report trust in businesses \citep{edelman2017}.

With regard to developing countries, the countries situated in the bottom twentieth percentile on corruption control are also the most fragile states \citep{Orre2008Mathisen}. States characterized by the weak central government and political instability also experience high levels of corruption \citep{FP2009}. These are often developing countries, which, according to The World Bank and the United Nations Office of Drugs and Crimes (UNODC), are estimated to lose USD 20-40 billion due to corrupt practices \citep{UNODC}. One of the direst statistics can be seen in the estimates of the African Union, which suggests that USD 150 billion (25\% of the continent's GDP) is lost every year due to financial malpractices \citep{TransparencyInternational2007}.

Corruption often occurs in countries rich in natural resources. When resource-rich countries have low performance in socio-economic development or a ``resource curse,'' corruption is seen as a critical factor in explaining it \citep{Mehlum2006,Robinson2006,Ross2001}. 



\subsection{Agency theory}

This paper uses agency theory as an analytical lens to identify the symptoms of contracts. Blockchain presents an alternative remedy to the symptom that has blighted the world economy. 

The relationship investigated in agency theory is one of the most commonly codified and oldest codes of social interaction. Agency theory revolves around the ``principal-agent'' relationship, where a ``principal'' hires an ``agent'' using a legal contract. As the ``principal'', delegates work to another, the ``agent'' \citep{Eisenhardt1985}, with the agent thereby acting on behalf of the principal in a particular decision problem domain \citep{Ross1973}.

The essence of agency theory is the study of principal-agent relationships such as lawyer-client, buyer-supplier, and employer-employee, to name a few \citep{HarrisRaviv1978}. The principal-agent literature intends to construct a blueprint for an optimal contract between principal and agent. Eisenhardt (1989) \citep{Eisenhardt1989} shows that the problem domain studied in the principal-agent relationship is where the principal and the agent have different preferences for what the goal of the relationship is, and how much risk they are willing to take.  The agent is, for example, compensated if one party prefers to whistle-blow when there are illegal practices in the relationship. Therefore, agency theory assumes that the principal and agent have different goals, known as a ``goal-conflict''.

Another aspect of the principal-agent relationship, ``moral hazard'', is defined as the case where an agent neglects contractual responsibilities. For example, moral hazard occurs if an employee uses company time to work on a personal project. The employee's tasks could be so complex that management fails to detect that what the employee is doing on company time. The employee (agent) violates the interests of the management (principal), and moral hazard occurs \citep{Eisenhardt1989,Holstrom1979}.

Another critical assumption in agency theory is known as information asymmetry, which emerges from the fact that a principal cannot monitor competencies and map out the agents' intentions beforehand. It also arises when knowledge of an agent's behavior is unclear and not deemed trustworthy. Eisenhardt (1989) \citep{Eisenhardt1989} notes that agency theory regards information as a commodity: Information has a cost and can be purchased, and organizations can invest in information systems to control opportunistic behavior from the agent.

As scholars such as Razavi and Iverson \citep{RazaviIverson2006} argued in 2006: Building a trusted network to share information increases interpersonal information exchange in supply chains. While this issue has been highlighted in agency theory, no optimal solutions have yet been provided. A possible explanation is that the symptom was identified before blockchain was conceived by a person or group under the pseudonym ``Satoshi Nakamoto'' in 2008 \citep{Nakamoto2008}. With the blockchain technology making significant progress over the past decade, models such as the MHC can be a cornerstone for increasing information sharing.

However, several studies argue that the effect of transparency on corruption is not absolute. In other words, transparency is a necessary but not sufficient condition to reduce corruption. In addition to access to information, you need an ability to process the information, and require the ability and incentives to act on the processed information.

\section{The Minimum Hybrid Contract (MHC)}

An MHC consists of a smart contract and a legal contract, where the legal contract is supplemented, not replaced, and effectively remains the same. The smart contract facilitates safe and transparent transactions, whose records are immutable, while the traditional contract -and the traditional institutions that ensure its validity- continues to take care of the required legal framework and disputes. The outcome of this synthesis of conventional legal structures and blockchain technology is the MHC, which aims to reduce corruption by increasing transparency, information sharing, and auditability. An additional benefit is a reduction in the cost of financial activity by replacing costly intermediaries such as accountants, banks, and clearinghouses with automated smart contracts.

The MHC is implementable using most of the blockchains currently available. We used Ethereum for running the smart contract because it is the most established of all Turing-complete smart contract blockchains. Ethereum also has the most intuitive Application Programming Interface (API), most active developer community, and very mature toolkit compared to the other options making it easier to adopt.

We define the following steps for the implementation of the MHC:

\theoremstyle{definition}\newtheorem{algorithm}{Algorithm}[section]

\begingroup
\makeatletter
\@beginparpenalty=10000
\makeatother

\begin{algorithm}
Pseudo-code for Figure \ref{fig_mhc-one-to-one-v2}. See \url{https://github.com/jakobsn/MHC} for implementation.%
\begin{enumerate}
    \item An MHC smart contract capable of referencing legal contracts and transferring funds is deployed.
    \item A legal contract $D$ is drafted with reference to the smart contract address.
    \item Actors sign the legal contract $D$, which is then archived/stored in digital format.
    \item The contract fingerprint $F_D$ is generated and added to the latest transaction block $B_k$ in the Ethereum blockchain. Including metadata such as the addresses for the actors.
    \item Actors in the contract can now send and receive transactions referring to the contract through the smart contract.
    \item In case of canceling the contract, both parties must use their personal Ethereum account to unsign (deactivate) the smart contract before the contract deactivates (see Algorithm 3.2).
\end{enumerate}
\label{abstract_algorithm}

\end{algorithm}
\endgroup

The public-private key pair is used in case of creating a digital identity for agents and principals. We illustrate the MHC architecture with the example of the Uport app from the Swiss municipality of Zug, which puts citizens in self-sovereign control of their identity \citep{Uport2017}. A public-private key-pair is generated for each citizen using the Uport app. Citizens can then share their public key with the municipality, and in response, the citizen is issued digital citizenship credentials.

The smart contract facilitates the transactions in the MHC. This makes the transactions instantly auditable because each transaction is indistinguishable from a receipt. Thus contractors and auditors can transparently audit the financial performance of the contract, and if there is a dispute, the legal contract is turned to for resolving it \citep{Grigg2004}.

\begin{algorithm}
Pseudo-code MHC implementation. The following are functions defined in the MHC running on the blockchain. See \url{https://github.com/jakobsn/MHC} for implementation.

\begin{itemize}
\item[] \textbf{function create\_contract}: Make a ContractStruct instance contract\_struct; Push the contract\_struct to the contract list; Map the participants to the contract\_struct; Record the contract creator and the contract id to the blockchain event log; Record that the party creating this contract also signs it.
\\
\item[] \textbf{function create\_contract\_signature}: Sign a contract registered on the blockchain; Once both parties have signed, the contract becomes active.
\\
\item[] \textbf{function create\_contract\_transfer}: Send funds through the MHC smart contract; Record the event of a contract transaction including the participants, amount sent and an invoice fingerprint.
\\
\item[] \textbf{function update\_contract\_unsign}: Unsign from a contract; Once both parties have unsigned, the contract deactivates.
\\
\item[] \textbf{function read\_contract}: Fetch the stored contract\_struct from the contracts list
\\
\item[] \textbf{function read\_contract\_ids}: Get the ids of the contracts related to an actor
\\
\item[] \textbf{function read\_is\_actor}: Check whether an actor is related to a specific contract id
\end{itemize}
\end{algorithm}

\begin{figure*}
\includegraphics[width=1\textwidth]{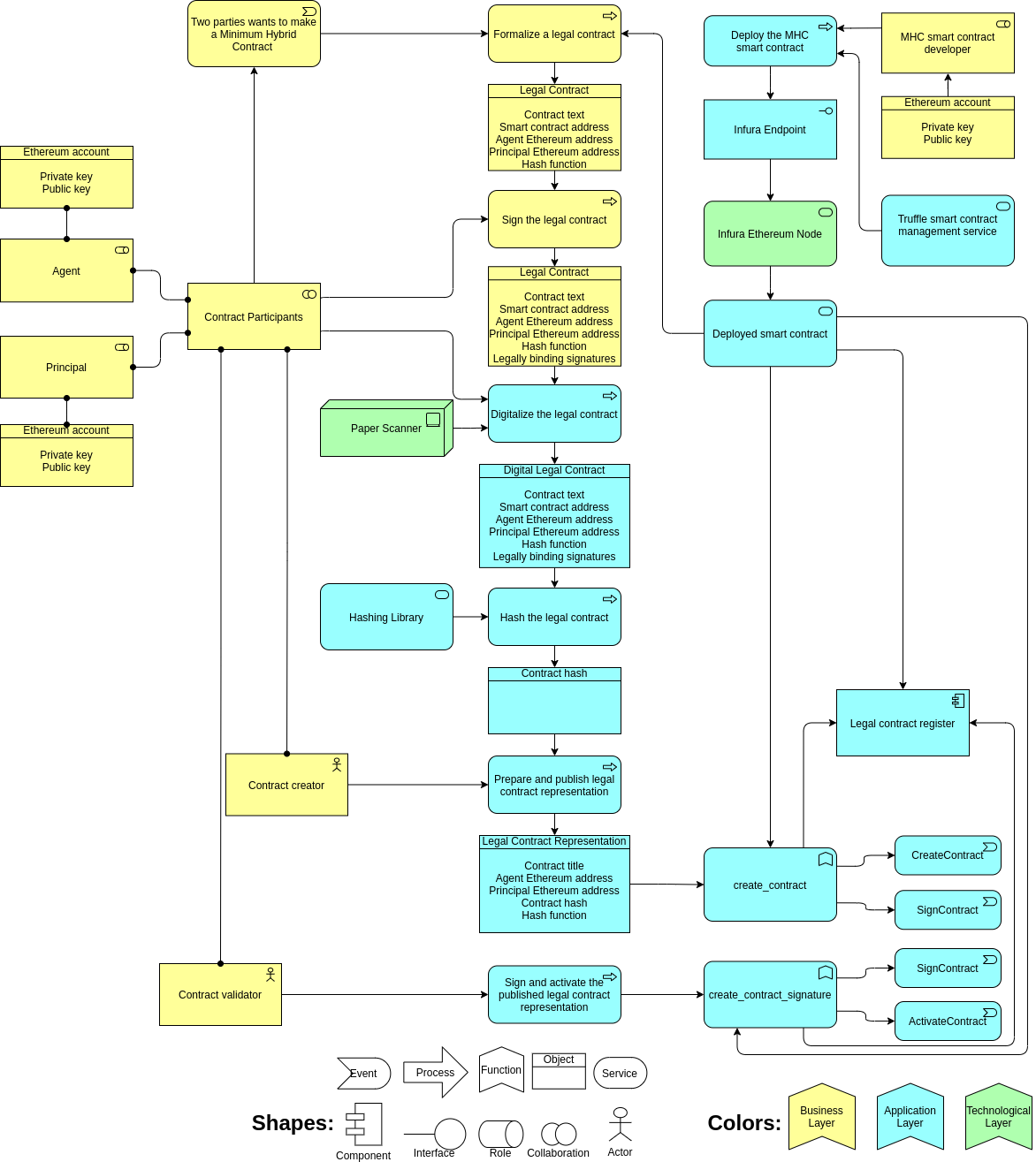}
\caption{The basis of an MHC, a one-to-one relationship between a legal and smart contract. }
\label{fig_mhc-one-to-one-v2}
\end{figure*}

In Figure \ref{fig_mhc-one-to-one-v2}, illustrates the relationship between a legal contract and the smart contract. The MHC smart contract is deployed for including the corresponding address in the legal contract. Further, one can imagine different implementations of deployed MHC smart contracts. The participants choose a suitable smart contract and define the legal contract. The next step for the participants is to commit to the legal contract by signing it. Once the legal contract is ready, it must be digitized. One can assume that the digital legal contract must be taken care of as if it was the paper-contract.  Since the only way to generate the digital fingerprint (hash) is by using the digital contract. 

The contract is now ready to be registered on the blockchain by using the create\_contract function. The registration includes the participants' addresses, the contract fingerprint, and the hashing method used for this contract. The fingerprint serves as a reference for the legal contract, and can only be generated from the unmodified legal contract. When a participant creates the contract representation, it must be signed by the opposing participant to become active. 

\begin{figure*}
\begin{center}
\includegraphics[width=1\textwidth]{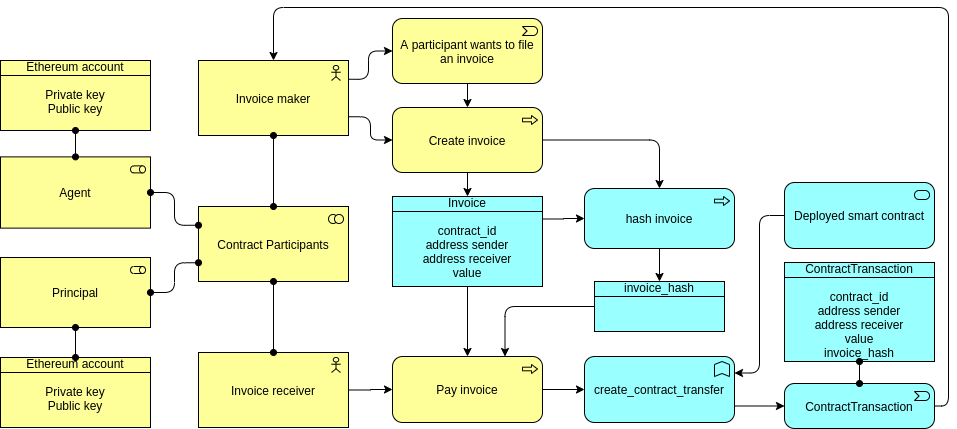}
\end{center}
\caption{A simple model of an MHC transaction using a smart contract and public-private key pairs.}
\label{fig_mhc-transaction}
\end{figure*}

Figure \ref{fig_mhc-transaction} shows how the MHC handles an invoice. Given the aforementioned one-to-one relationships, the transactions in the smart contract have legal status. When a participant wants to make an invoice for a certain amount of cryptocurrency, the first step is to create a digital invoice. The MHC smart contract can hash the invoice in the same fashion as the legal contract to provide a reference to the invoice. A receiver receives the invoice and a hash of it and promptly signs it with their private key to authenticate the broadcast of the corresponding transaction.  The smart contract records the contract reference, invoice hash, sender, receiver, and amount. When the blockchain approves the transaction, the invoice maker receives the cryptocurrency. 

\begin{figure*}
\includegraphics[width=\textwidth]{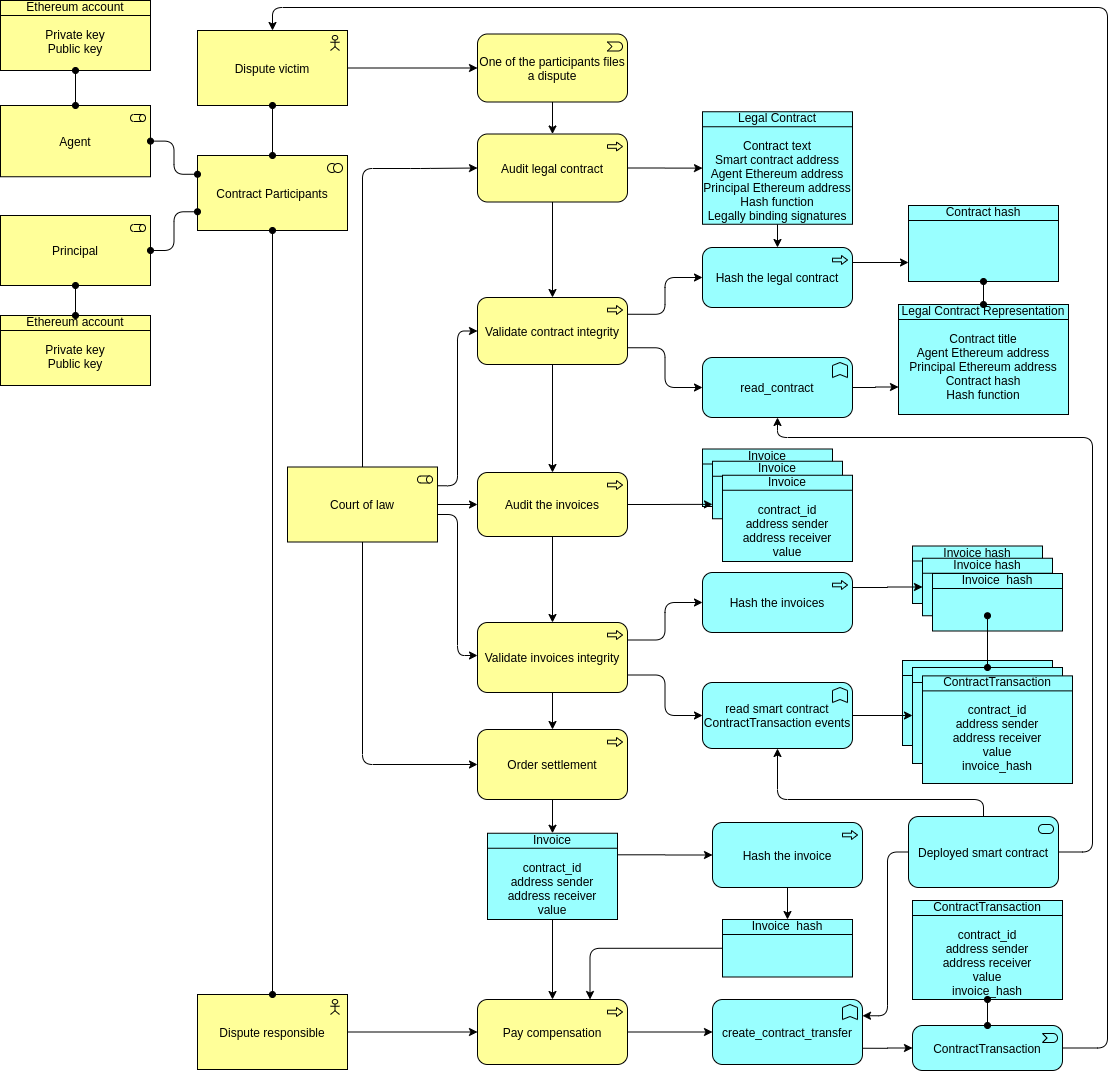}
\caption{A simple model of the most relevant elements of a dispute in an MHC.}
\label{fig_mhc-dispute}
\end{figure*}

The MHC can transparently unveil illegal and fraudulent behavior by auditing transactions from smart contracts. When using these transactions to report such disputes to a court for investigation, the legal contract, and actions taken in the contractual relationship such as transactions that are being openly auditable. If a court orders a financial settlement, a smart contract cannot force a reversal of transaction because immutability is a central capability of smart contracts. Thus a new transaction has to be undertaken. Figure \ref{fig_mhc-dispute} shows the dispute process under the MHC.

\subsection{Experimental results}

The MHC smart contract has been deployed to the Ethereum Main Net to demonstrate our use case . We used the MHC architecture and the Ethereum programming language (Solidity) \citep{solidity} for the touring-complete Ethereum Virtual Machine (EVM) \citep{evm}. The smart contract contains the methods from the pseudo-code in Algorithm \ref{abstract_algorithm}. To upload the smart contract to Ethereum, the Truffle Command Line Interface (CLI) \citep{truffle} was used in addition to Ethereum cryptocurrency ETH to pay transaction fees. The goal of this experiment was to show how we achieve transparency, immutability, and auditability using the MHC. 

The implementation of the MHC smart contract now runs openly on the Ethereum blockchain on this Ethereum address: \linebreak \texttt{0xC0e817d3e3D79085175BCD83BD899A3d06BC1C8A}. Anyone can use the smart contract in that address to immutably link a legal and smart contract. See the README.md file in the following MHC Git repository for more details: \url{https://github.com/jakobsn/MHC}. 

Through our experiments we create a legal contract representation providing a unique fingerprint hash, this hash can only be generated with a specific file such as a digital legal contract. The hash is immutable because once it has been recorded on the blockchain it can never be changed. Once both participants have signed the contract, the contract activates and an event is recorded to the contracts event log. Events are recorded with a reference to their contract instance whenever participants perform an action, such as a creating a contract instance, making a contract transaction or unsigning from a contract. The events are linked to their transaction, including additional information such as block height, timestamp, sender and receiver. Combining the event log with the stored contract representations, the system can be audited at any time and provides complete transparency.

\subsection{Applicability}

Legislators have already begun supplementing contracts with smart contracts. Delaware has adopted legislation authorizing smart contract ledgers for state records and regulatory functions such as tracking liens and corporate shares \citep{Roberts2017}. Arizona introduced a law declaring that blockchain-based digital signatures are legally enforceable \citep{Higgins2017}, and Vermont \citep{Vermont2018} has declared that blockchain-based evidence is admissible in court. This Suggests that blockchain, irrespective of the MHC, is increasingly becoming an established standard for transferring value over the internet.

Legal scholar Werbach has argued that smart contracts are capable of serving in legal compliance with and directly shape compliance with legal obligations if integrated correctly \citep{Werbach2018}. Therefore the architecture of the MHC is outlined as it is the most revolutionary and conservative. That is because this paper wants to make the MHC an iterative suggestion and keep the legal contract intact, thus making the MHC backward compatible with all legal contracts. In addition to that, the MHC leverages the transparent and immutable properties of the smart contract as it conducts the financial transactions. Because of its simplicity, this paper argues that that the MHC is the hybrid architecture regulators most likely agrees with initially.

\section{Discussion}
The MHC offers transparency and immutability for financial transactions, and the receipt for a transaction is indistinguishable from the transaction itself. There is zero opportunity to fake receipts, and thus receipt fraud is theoretically impossible. Using MHC's immutable and transparent transactions for cost-accounting can reduce the risk of illicit activity because of superior reliability and cost-accounting.

Financial information is usually costly to obtain and flawed. The price of information decreases when using the MHC, as there is no longer a need to employ an auditing firm to review financial records, and there is no need to trust the auditor and the issuer and safe keeper of the receipt. The principal can audit the records in the MHC smart contract to prove which past transactions that were signed using the agent's public key-pair.

In the information asymmetry paradox by Arrow (1962) \citep{Arrow1980}, it says that nobody can know what information they buy before they have it. Therefore a buyer of information cannot know exactly what he is buying. The MHC has the capability to prove that a particular individual or organization conducted a immutable and transparent transaction. Information asymmetries also arise when the knowledge of an agents behavior is unclear. Therefore making financial behavior instantly auditable will help mitigate the information asymmetry paradox. As in the case of complete and incomplete information, the MHC gives the principal the capability to have full details of all financial transactions conducted by the agent. The principal can use the MHC for budgeting and complete information on cost-accounting and complete information in auditing \citep{Eisenhardt1989}. 

In the case of a principal having incomplete information, Eisenhardt (1985) \citep{Eisenhardt1985} notes that monitoring behavior is a solution. Here the principal can purchase information and reward them. As an MHC provides a cost transaction system built into the transactions itself, this creates an environment where the information is not necessarily purchased but audited of the blockchain transaction ledger directly on the internet.

Access to information and transparency is normatively considered a human right in democratic societies, as it is assumed to constitute a fundamental right to be informed about government actions and the reasons for them. In practical terms, transparency is essential for human development as it provides incentives for inclusiveness and redistribution \citep{Stiglitz1999}.

To minimize moral hazard, principals incurring monitoring expenditures, for example, the cost of using external auditors for scrutinizing financial statements. On the other hand, agents can incur costs when purchasing an internal audit to signal to the principal that they are acting inconsistently with the legal contract (Adams 1994). Wallace (1980) argues that the agent's compensation reflects what a principal expends for monitoring an agent's actions.

It is crucial to consider an MHC not exclusively as a financial tool but also as a governing tool. For example, in the 1990s, Uganda surveys showed that only 13\% of education grants reached schools, and the rest was embezzled in the process by the local government. When the Ugandan government(Principal) started publishing every monthly grant in district newspapers, there was a substantial effect on grants received by schools (agents). Surveyed in 2001, over 80\% of the grants reached a school (agent), on average. Thus the impact of access to information about the grants and thus agent's rewards were positive \citep{ReinikkaSvensoon2005}.


Considering that transactions in the MHC can be audited instantly and are immutable, all financial actions of agents are transparent to the principal. Thus, the MHC is a transparency and surveillance mechanism for financial information such as auditing, accounting, cost accounting, and budgeting.

The MHC reduces the cost of enforcing financial transactions and auditing in contracts and defines property rights of money using public-private key pairs for addresses and authenticating transactions. Practises for accounting and disclosure become automated and removes potentially corrupt personnel and the capability to tamper with records. Regulators can trust more in market integrity as the financial information they receive about the market is more accurate and cannot be tampered with by corrupt actors.

To gradually test and implement the MHC financial regulators are recommended to use ``safe harbors'' where specific activities are excluded from legal enforcement and ``regulatory sandboxes,'' which is similar but regulated in time and scale. Using these strategies regulators can collaborate effectively with actors using MHCs \citep{CDA1996}.

\section{Conclusion}
In light of how current contracting practices and legislators globally are all affected by systemic corruption at some level, this paper has developed an architecture for the Minimum Hybrid Contract (MHC). The MHC is a practical and evolutionary step for contracting using smart contracts as supplements to legal agreements for financial transactions. Agency theory, which focused on contractual relationships, has been used to discuss the capabilities and implications of the MHC. Finally, this paper focuses on how MHCs can affect agency theory issues as it increases information sharing, automates cost accounting, and can potentially mitigate moral hazard because the financial transactions in the MHCs and immutable transactions enables more trustworthy and cost-effective auditing.

The MHC does not place trust for financial activity in individuals and central institutions, but instead distributes it across the user base and replacing potentially questionable central authorities by communities of peers in the form of peer-to-peer networks. Depending on and trusting a third-party for financial transactions is no longer needed. In the democratized context offered by blockchain, corporations and superpowers cannot unilaterally defy the community and break the rules of the system; thereby, the MHC can create a more trustworthy system \citep{Sun2016,ScottsZachariadis2012}.

MHC affects information asymmetries, moral hazard, and monitoring costs in agency theory. In light of the MHC, agency theory is most affected by its transparent and immutable properties. Eased sharing and increased transparency of financial information lower the cost of sharing information and decreases information asymmetries. The MHC reduces the monitoring cost of the principal as it reduces the cost of obtaining financial information and conducting audits. The MHC provides a similar network as Razavi and Iverson (2006) \citep{RazaviIverson2006} calls out for in supply chains to exchange interpersonal information, but specifically for financial information. The MHC's transparent and immutable properties make it a deterrent against financial crime, potentially reducing moral hazard.

Blockchain regulation strategies are needed to set forth a pathway for adopting the MHC. If blockchain is made illegal, the MHC cannot use a smart contract as a supplement to a legal contract. The recommended strategy for pioneering regulators is to implement safe harbors and regulatory sandboxes where specific blockchain-related activities are excluded from legal obligations.

To utilize the MHC for societal well-being finding a golden middle way between private and transparent transactions is essential, and striking a balance must between transparency and the basic human need for privacy and secrets \citep{Harari2014} is essential. ``Satoshi Nakamoto'' (2008) \citep{Nakamoto2008} invented a way to conduct open, borderless, censorship-resistant, and neutral transactions without having to trust any corruptible central authority is a game-changer for making contracts more trustworthy, and will most definitely affect all of finance. Since blockchain transactions are transparent, with immutable properties, it is also more trustworthy and makes auditing more cost-effective. Ultimately it succeeds in mitigating moral hazard, information asymmetries, and monitoring costs emphasized by agency theory. Because of regulatory uncertainty around blockchain, the authors recommend regulatory strategies such as sandboxes and safe harbors for implementing the MHC cautiously.

\section*{Acknowledgements}

The authors would like to thank Emil Sina Emami for his helpful feedback and suggestions. The authors would also like to thank the reviewers from Evaluation and Assessment in Software Engineering (EASE), 2020 Trondheim, Norway, for their constructive feedback.

\bibliographystyle{unsrtnat}
\bibliography{references.bib}

\end{document}